\documentclass{PoS}
\newcommand{\signhw}{\mbox{sign}(H_w) }
\title{Electric polarizability of hadrons with overlap fermions on multi-GPUs}

\ShortTitle{Electric polarizability of hadrons with overlap fermions on multi-GPUs}

\author{\speaker{Michael Lujan},~~Andrei Alexandru,~and~Frank Lee\\
        The George Washington University, Washington DC, USA\\
        E-mail: \email{mlujan@gwmail.gwu.edu},~~ \email{aalexan@gwu.edu},~~ \email{fxlee@gwu.edu}}

\abstract{Electric polarizability is an important parameter for the internal structure of hadrons. Previous studies of polarizabilities have been done at relatively heavy pion masses, leaving the chiral region largely unexplored. In this report, we use overlap fermions which are known to be computationally demanding to properly capture the chiral dynamics. We present an implementation strategy to construct overlap on multi-GPUs. We find that our GPU code has an equivalent of $\sim 30$ CPU cores to 1 GPU. We also present preliminary results for the polarizability of the neutral pion.}

\FullConference{ The XXIX International Symposium on Lattice Field Theory - Lattice 2011\\
July 10-16, 2011\\
Squaw Valley, Lake Tahoe, California}

\begin{document}

\section{Introduction}
Electric polarizability is an important characterization of the internal structure of hadrons. However, current lattice QCD calculations use pion masses $m_{\pi} >$ 300 MeV which is a factor of 2 greater than the physical mass. Yet, at relatively low pion masses that lie above the physical point, interesting physics is still predicted to occur. For example, when $m_{\pi}$ is smaller than $m_{\Delta} - m_n$, the polarizability is expected to change substantially \cite{Hildebrandt:2005iw}. 

Simulations performed near the physical point are challenging. The effects of chiral symmetry breaking become important at lower masses. Furthermore, exceptional configurations appear more frequently in the simulations \cite{Alexandru:2009id}, making the inversion of the Dirac operator more problematic.  To overcome these difficulties, we use the overlap operator \cite{Neuberger:2001ig} which preserves exact chiral symmetry on the lattice and does not suffer from exceptional configurations. However, the overlap operator is computationally expensive. Due to this we implement it on Graphic Processing Units (GPUs). 

In recent years, the usage of GPUs has increased in the lattice community \cite{Alexandru:2011ee}, \cite{Alexandru:2011sc}, \cite{Clark:2009wm}, \cite{Egri:2006zm}, \cite{Babich:2010mu}. They are known to outperform CPUs by large factors.  However, one drawback is that they have relatively small memory compared to CPUs. The overlap operator is not only numerically demanding but it is also memory intensive, forcing us to implement overlap using multi-GPU architectures.

The outline of the paper is as follows: In section 2, we introduce the overlap formalism and the method we use to approximate it. Section 3 describes the implementation of overlap on GPUs and how we compute overlap quark propagators; here, we also address the issue of memory constraints. In section 4, we describe the methodology of how we compute the polarizability. In particular, the background field method \cite{Alexandru:2008sj} is discussed. Lastly, we present our results on the neutral pion.

\section{Overlap Formalism}
The Dirac overlap operator is defined as
\begin{equation}
D_{ov} = 1 + \gamma_5~ \mbox{sign}(H_w),
\label{eqn:ov}
\end{equation}
where $H_w \equiv \gamma_{5}D_w$ and $D_w$ is the Wilson Dirac operator. It is numerically not feasible to compute $\signhw$ exactly. One must use numerical methods to approximate the matrix sign function.  There are 2 commonly used approximations: polynomial \cite{Giusti:2002sm} and rational \cite{Chiu:2003ub} approximations. In this study, we use the polynomial approximation.  One constructs a polynomial $P(x) =( p_0 + p_1x +p_2 x^2 +..p_n x^n)$ to approximate $x^{-1/2}$.  The matrix sign function is then approximated as $\mbox{sign}(H_w) \approx Q P(Q^2)$, where $Q = H_w/||H_w||$.  

The order of the polynomial can be estimated using the empirical formula  \cite{Giusti:2002sm} 
\begin{equation}
\delta = A e^{-b n \sqrt{\epsilon}},
\label{eqn:error}
\end{equation}
where $A = 0.41$, $b=2.1$. The parameter $\delta$ is the error of the approximation {\it{i.e.}} $\delta = \left| 1 - \sqrt{x}P(x)\right|$. The parameter $\sqrt{\epsilon}$ defines an interval, $[-\sqrt{\epsilon},\sqrt{\epsilon}]$, around zero where the approximation breaks down. Because we want the approximation to be valid over the whole spectrum of $H_w$, $\sqrt{\epsilon}$ needs to be less than the smallest eigenvalue of $H_w$. In our runs, we find $\lambda_{min} \sim 10^{-4}$. The corresponding polynomial order is $\sim 10^5$ for $\delta = 10^{-10}$. However, this is impractical.

\begin{figure}[htp]
\begin{center}
\includegraphics[width= 2.8in]{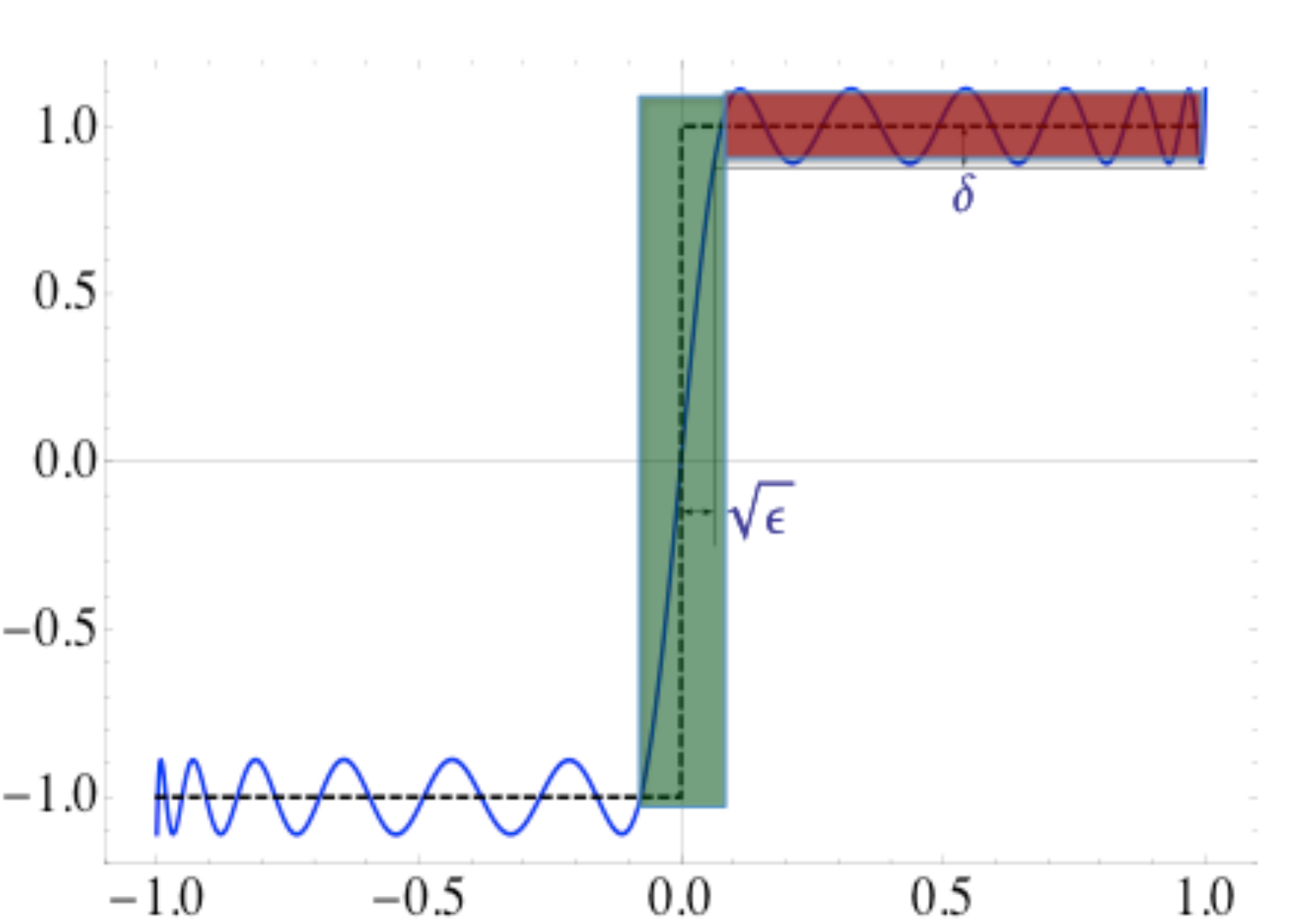} 
\caption{Polynomial approximation $xP(x^2)$ for the sign function with $\delta = 0.1$, $n=13$, and $\sqrt{\epsilon} = 0.06$. This is a schematic diagram of how the sign function approximation is divided into 2 parts: small space and large space. The green/vertical and red/horizontal region corresponds to the small space and large space, respectively.}
\label{sign}
\end{center}
\end{figure}

The general approach to this problem is to divide the approximation into 2 regions which we call the small space and large space. The small space is computed exactly by calculating a small spectrum of $Q$ around zero. The remaining large space is approximated by $\mbox{sign}(H_w) \approx Q P\left(Q^2\right)$ with a significantly smaller polynomial order. This makes the calculation more feasible. Figure \ref{sign} shows a small order approximation and the breakdown of the 2 regions.

We compute the quark propagators using adaptive conjugate gradient (CG) methods. For small quark masses, the convergence is very slow. Therefore, we use deflation to help speed up the calculation.  This consists of computing the low lying spectrum of the overlap operator.  For pion masses on the order of 200 MeV, deflation was shown to increase the convergence rate by a factor of 3-4 \cite{Li:2010pw}.  The process of obtaining the eigenmodes for $D_{ov}$ is similar to the computation of the eigenmodes of $H_w$ that is used for the small space calculation.  We now describe how to implement these eigensystem and propagator calculations on GPUs.

\section{Overlap implementation on multi-GPUs}
GPUs have been very successful thus far in lattice QCD calculations. They have good floating point performance and large memory bandwidth. However, a drawback of GPUs is their fixed memory size, ranging from 1-6 GB per GPU. Because the CPUs are bottlenecked by communication over the PCI bus, it is advantageous, though not always possible, to keep all necessary data on the GPU memory. 

In this study, we used a lattice size of $24^3 \times 64$. The calculations were done on a GPU cluster consisting of 6GB of memory per GPU.  On a single GPU, we can store 36 vectors.  We will show that due to memory constraints, we cannot compute overlap propagators on a single GPU, forcing us to use multi-GPU architectures. 

To compute an overlap quark propagator, there are 2 main routines that need to be implemented: an eigensystem solver and a multi-mass inverter. The eigensystem solver is used to compute the eigensystems of $H_w$ and $D_{ov}$.  We discuss these 2 routines in the following sections.

\subsection{Eigensystem solver}
The practical implementation of the overlap operator requires the calculation of the lowest lying eigenmodes of $H_w$.  In order for one to use deflation, we also need to compute the lowest eigenmodes of $D_{ov}$.  To calculate these eigenmodes, we use implicitly restarted Arnoldi factorization \cite{Alexandru:2011sc}.  In this algorithm we build a subspace of the size $2.5 l$, where $l$ is the number of desired eigenmodes. In most cases, $l$ is too large so that the required memory cannot fit into a single GPU. 

\begin{figure}[htbp]
\begin{minipage}[b]{0.5\linewidth}
\centering
\includegraphics[width= 2.8in]{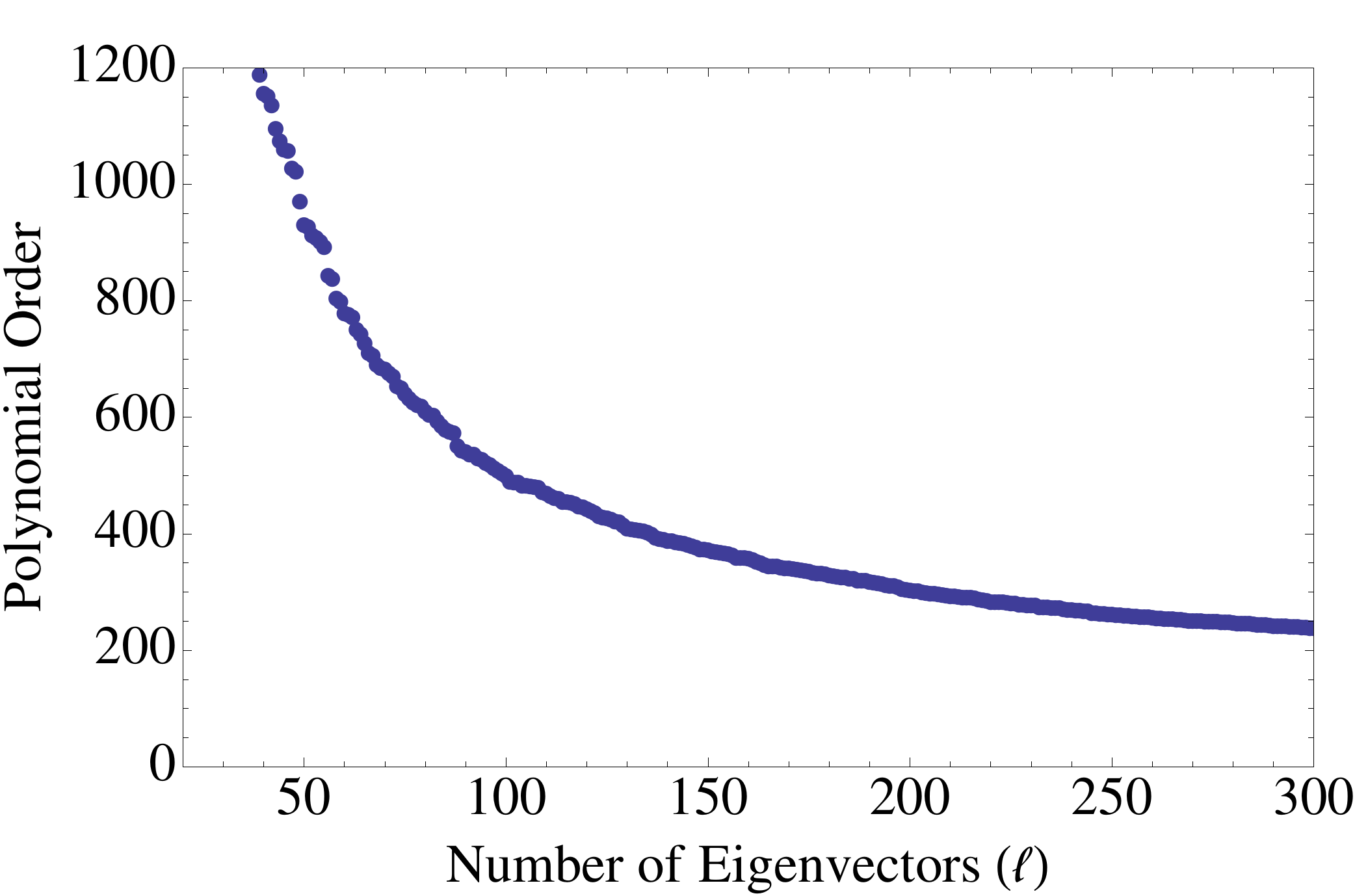} 
\caption{Polynomial order to approximate the sign function as a function of $l$.}
\label{hw}
\end{minipage}
\hspace{0.2 in}
\begin{minipage}[b]{0.5\linewidth}
\centering
\includegraphics[width= 2.8in]{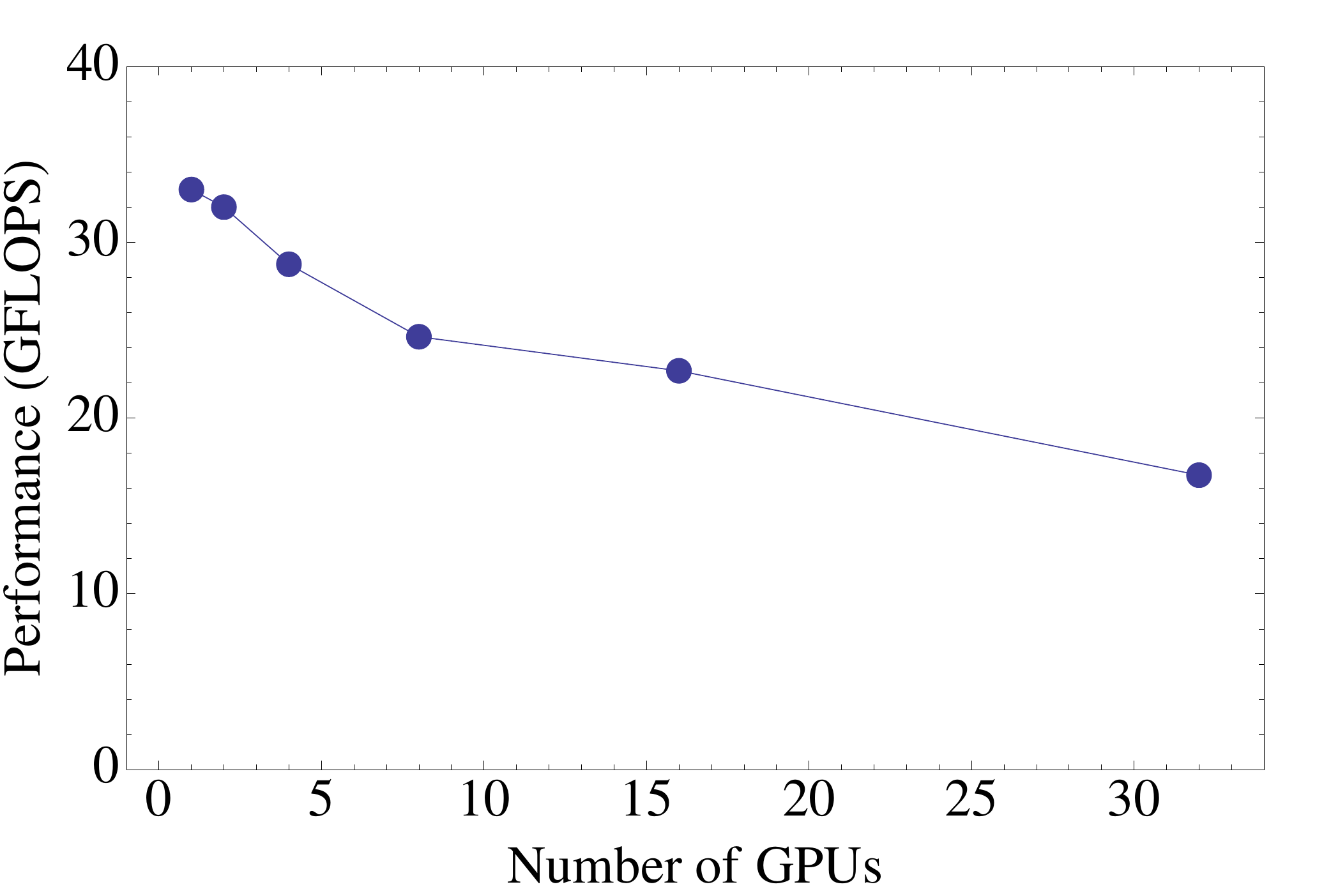}
\caption{Strong scaling of the double precision dslash on $24^3 \times 64$ lattice.}
\label{dslash}
\end{minipage}
\end{figure}

For the $H_w$ eigensystem the choice of $l$ is dictated by the spectrum of its eigenvalues.  It has been shown \cite{Alexandru:2011sc} that the eigenvalue distribution of  $H_w$ varies little from configuration to configuration.  Therefore, one can calculate a $H_w$ eigensystem from one configuration and use eqn. \ref{eqn:error} to compute the corresponding polynomial order to determine an optimal choice of $l$. We note that for a given $\delta$ there is an inverse relation between $n$ and $\lambda_l$. A larger $\lambda_l$ corresponds to a smaller $n$, which is desirable. It has been shown \cite{Li:2010pw} that using smeared operators increases the magnitude of the eigenvalues for the low lying eigenmodes. In this study, we smear  the gauge links 3 times using nhyp smearing  \cite{PhysRevD.69.054501}.  In fig. \ref{hw} we plot a small spectrum of  $H_w$ as a function of the number of eigenvalues for a single configuration. Going from $l = 100$ to 200, $n$ is reduced by about a factor of 2. However, going from 200 to 300 there only is a 15\% reduction. Moreover, $l=300$ requires 1.5 times more memory, we therefore choose $l=200$. The advantage of using smaller values of $l$ is discussed below.

With $l=200$ a minimum of 16 GPUs are needed in order to hold this amount of data.  One should always use the least amount of GPUs required to do the calculation; this is due to the scaling of the dslash operator.  Recall that the overlap calculation requires a large amount of dslash multiplications. Our timings indicate that approximately 70\% of the overlap calculation is spent in the dslash routine. The scaling of dslash directly affects the scaling of overlap. Figure \ref{dslash} shows the dslash scaling from 1 to 32 GPUs.  From the plot we see that it is beneficial to use the least amount of GPUs.

Chebyshev acceleration \cite{Neff:2001zr} has also been implemented in this calculation. This serves 2 purposes.  First, it speeds up the convergence of the $H_w$ eigensystem. We use a Chebyshev polynomial of order 100. The algorithm usually converges in 1 iteration.  Second, it offers an alternative when GPU memory alone is insufficient.  By using the Chebyshev acceleration in conjunction with the Arnoldi eigensolver, we can utilize CPU memory. However, in the case of $H_w$ this may impact the performance of the code substantially as explained below.

In simulations where GPU memory is not sufficient, we implemented a mixed code to utilize both CPU and GPU memory. CPU memory is generally much larger than GPU memory; one can store all the $H_w$ eigenvectors on the CPU. If the eigensystem solver uses no Chebyshev acceleration, we would have to transport each vector to and from the CPU memory after a {\it{single}} dslash multiplication. The overhead associated with transporting the vectors over the PCI bus would be much greater than the time spent in the calculation which would make the code very inefficient. However, by using Chebyshev acceleration one needs to perform a few hundred dslash multiplications for each vector transported. This hides the communication overhead. The drawback is that all vector routines, such as vector additions and scalar products, are computed on the CPU.  This can impact the performance of the code substantially. In particular, the larger value of $l$ will have a worse performance because the number of dslash operations (done on GPU) scales as $l$ while the number of vector routines (done on CPU) scale like $l^2$ in the Arnoldi algorithm. 
 
The GPU cluster used for our runs has 6GB of memory per GPU  with QDR Infiniband. We use 16 GPUs in order to fit all data into  GPU memory. To compare the GPU performance with CPU codes, we run similar calculations on a Cray XT-5 machine. Since the scaling of the CPU codes is poorer than our GPU codes, we run our codes on 256 cores which is the minimum required to complete our tests in the time limit imposed by the scheduling system.
 
In our runs with 16 GPUs and $l=200$ the all GPU code takes 510 seconds, while the mixed code takes 1600 seconds. Our all CPU code took 1100 seconds.  Compared to the all GPU code and the machines we ran on, we find that there is an equivalent of about 34 CPU cores to 1 GPU. For the mixed algorithm there is an equivalent of 11 CPU cores to 1 GPU.
  
Once the $H_w$ eigenvectors have been computed, we can obtain the eigensystems for $D_{ov}$ which will be used for deflation to help speed up the propagator calculations.  We again use the Arnoldi eigensolver to compute the eigensystem in a similar manner as was done with $H_w$ which includes using Chebyshev acceleration.  A mixed approach using CPU and GPU memory can also be used for computing the eigensystem of $D_{ov}$ if GPU memory is limited. Because $D_{ov}$ requires all of the $H_w$ eigenvectors for each matrix multiplication, we keep all $H_w$ eigenvectors on the GPU and all $D_{ov}$ eigenvectors on the CPU.  In contrast to what was seen in the case of $H_w$, there is only a 40\% slowdown going from an all GPU code to the mixed code. We used a Chebyshev order of 12 for these runs. The calculation of the 100 lowest eigenmodes for the overlap operator required a total of 4.8 hours on our all GPU code and 6.7 hours for our mixed code. The main reason for the increase of relative performance of $D_{ov}$, in comparison to $H_w,$ is due to the fact that $D_{ov}$ has to perform many more dslash operations for each Arnoldi iteration. For the mixed code one should notice that the calculation of eigenmodes for $D_{ov}$ is much more expensive than for $H_w$.  This means that the factor of 3 in performance lost in the mixed $H_w$ code is not substantial when considering the whole calculation of eigensystems. Again, we can compare our GPU code to an all CPU calculation. The all CPU required 11.2 hours on 256 cores for the eigensystem calculation of $D_{ov}$. Thus, there is a GPU equivalent count of 37 and 27 for the all GPU and mixed code, respectively.

\subsection{Quark propagator calculation}
To compute $D_{ov}^{-1}~ \psi$ with a given precision we implemented an a adaptive multi-mass CG method \cite{Cundy:2004pza}.  The precision of $D_{ov}$ can be tuned by changing the order of the polynomial $n$.  As the CG process converges less precision is required at each step. This allows us to change the order of the polynomial accordingly. Furthermore, at some point the precision required can be achieved using single precision arithmetic. This produces an even better performance. For a quark mass corresponding to $m_{\pi} \approx 200$ MeV and a residue of $10^{-8}$, the adaptive CG took 0.8 hours, while the non-adaptive CG required 1.6 hours.

The most expensive procedure for computing an overlap quark propagator is the calculation of eigensystems for $D_{ov}$.  Altogether, a complete quark propagator calculation for this work takes roughly 6 hours using 16 GPUs with the all GPU code. We apply this overlap implementation to compute the polarizability of hadrons.  We will need 5 quark propagators per configuration in order to calculate the polarizability. We now turn our attention to the methodology for extracting the polarizability. 

\section{Methodology and Results}
We use the background field method to introduce a constant electric field onto the lattice. The basic formulation is to modify the covariant derivative in the following way

\begin{equation}
D_{\mu} = \partial_{\mu} -iG_{\mu} - iq A_{\mu},
\end{equation}
where $q$ is the electric charge. $G_{\mu}$ and $A_{\mu}$ are the gluon and photon fields, respectively. $A_{\mu}$ has the effect that it multiplies all the gauge links by an extra $U(1)$ phase factor {\it{i.e.}} $U_{\mu} \rightarrow e^{-iqA_{\mu}}U_{\mu}$. 

In this work, we focus on the neutral pion polarizability.  One method to extract the energies for the neutral pion is by looking at the ratio of correlation functions.  This is given by
\begin{equation}
R(t) = \frac{G_E(t)}{G_0(t)} \rightarrow \frac{e^{-(m+\Delta m) t}}{e^{-mt}} = e^{-\Delta m t},
\end{equation}
where $G_E$ and $G_0$ are the correlation functions with and without the electric field, respectively. $\Delta m$ is the desired energy shift of the hadron. Boundary conditions play an important role for the background field method \cite{Alexandru:2010dx}. In this study, we use Dirichlet boundary conditions both in the direction of the applied electric field and time. The boundary condition in the spatial direction has the effect of producing a non-zero total momentum for the system. In order to extract the polarizability, one needs to first subtract off the additional energy based on the fact that the pion is moving.

The calculation of $R(t)$ is relatively expensive.  To obtain a better signal for $G_E(t)$ we compute propagators with both positive and negative values of the electric field. Furthermore, for each non-zero value of the electric field we need to compute $u$ and $d$ propagators separately. This is because the $u$ and $d$ quarks couple differently to the external field due to their different electric charges. Thus, we need to compute 4 quark propagators to calculate $G_E(t)$ and 1 to calculate $G_0(t)$, a total of 5 propagators per configuration.  This requires 5 different calculations of the $H_w$ and $D_{ov}$ eigensystems.

We use 40 configurations of the 2+1 domain wall fermion gauge configurations \cite{Mawhinney:2009jy} on $24^3 \times 64$ lattices with $a^{-1} = 1.73(3)$ GeV and a pseudoscalar sea mass $\approx 330$ MeV. Each configuration was smeared 3 times using nhyp smearing.  We used 4 different source positions, equalling a total of 160 propagators for each value of the external field. Adding a new source is relatively inexpensive since the $H_w$ and $D_{ov}$ eigensystems do not have to be recomputed. A total of 7 masses were used, with the lowest mass roughly 240 MeV. 

\begin{figure}[htp]
\begin{center}
\includegraphics[width= 3.1in]{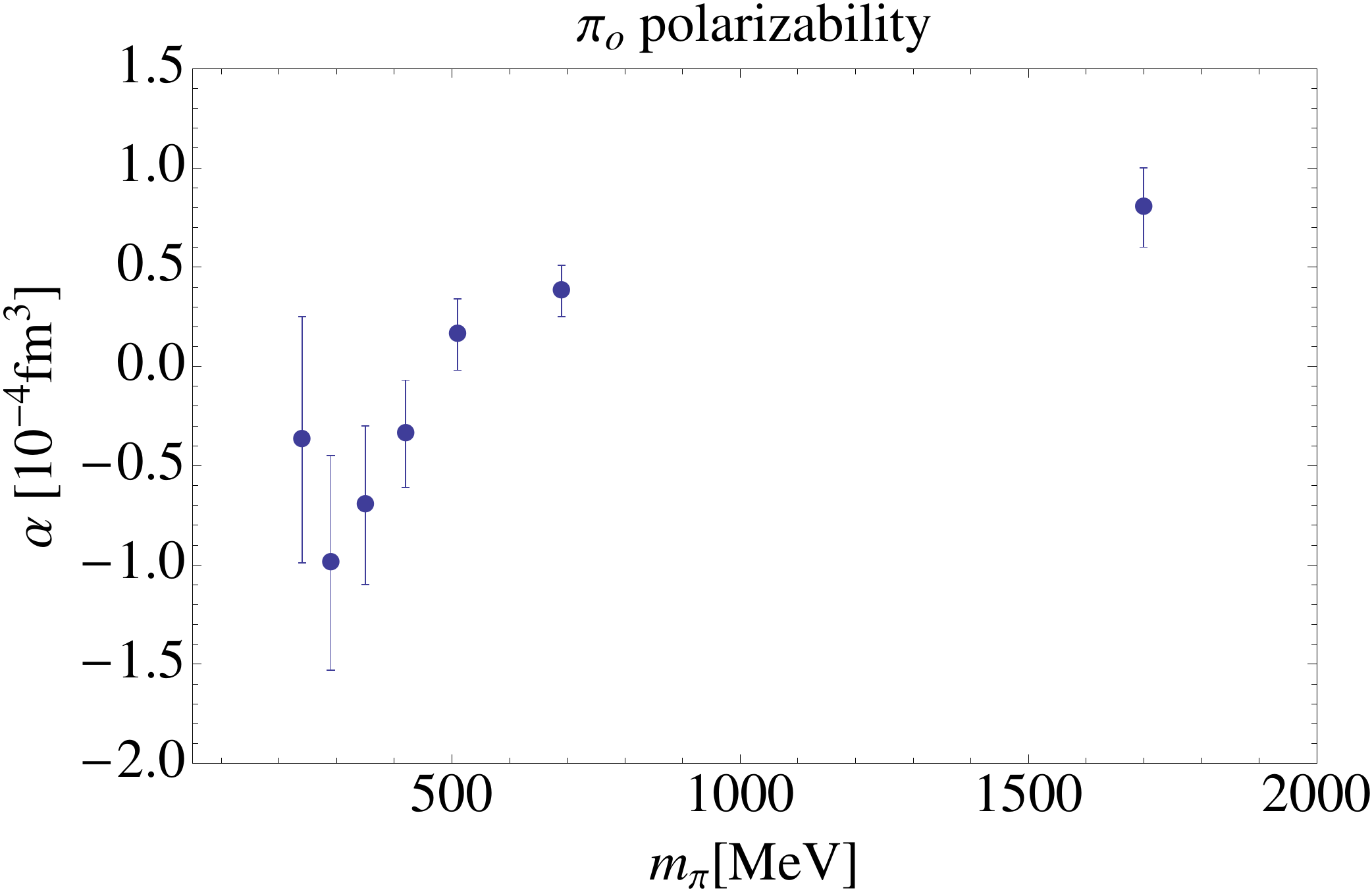} 
\caption{$\pi_0$ polarizability.}
\label{plot:polar}
\end{center}
\end{figure}

Figure \ref{plot:polar} shows a plot of the extracted polarizabilities for the neutral pion. We see that at $m_{\pi} \approx 500$ MeV the polarizability changes sign.  This is consistent with other studies \cite{Alexandru:2009id}.  Our error bars are rather large due to small statistics used. We are currently generating 4 times the statistics to get a cleaner signal at lower masses. 

\section{Conclusion}
The calculation with overlap fermions is very computationally demanding. However, we have constructed an efficient implementation of the overlap operator on GPUs. We presented a strategy to utilize both CPU and GPU memory if GPU memory alone is insufficient to hold all of the required data. Our simulations with the all GPU code indicate that a single GPU performance is equivalent to roughly $\sim 30$ CPU cores on the machines used. We have also presented preliminary results of the neutral pion polarizability using overlap fermions. We are currently in the process of generating a factor of 4 in statistics in order to reduce our error bars. 

\bibliography{my-references-2011_lattice}
\end{document}